\documentclass[submission, Proceedings]{SciPost}

\hypersetup{
    colorlinks,
    linkcolor={red!50!black},
    citecolor={blue!50!black},
    urlcolor={blue!80!black}
}
\usepackage{ifthen}
\newboolean{uprightparticles}
\setboolean{uprightparticles}{true} %Set true for upright particle symbols
\usepackage{xspace}
\usepackage{upgreek}

\def\MagUp {\mbox{\em Mag\kern -0.05em Up}\xspace}

\ifthenelse{\boolean{uprightparticles}}%
{
 
 \def\Pgamma      {\ensuremath{\upgamma}\xspace}

 \def\Peta        {\ensuremath{\upeta}\xspace}

 \def\Pmu         {\ensuremath{\upmu}\xspace}
 \def\Pnu         {\ensuremath{\upnu}\xspace}
 
 \def\Ppi         {\ensuremath{\uppi}\xspace}

 \def\Pchi        {\ensuremath{\upchi}\xspace}
 \def\Ppsi        {\ensuremath{\uppsi}\xspace}

 \def\PDelta      {\ensuremath{\Delta}\xspace}
 \def\PXi      {\ensuremath{\Xi}\xspace}
 \def\PLambda      {\ensuremath{\Lambda}\xspace}
 \def\PSigma      {\ensuremath{\Sigma}\xspace}
 \def\POmega      {\ensuremath{\Omega}\xspace}
 \def\PUpsilon      {\ensuremath{\Upsilon}\xspace}

 \def\PB      {\ensuremath{\mathit{B}}\xspace}
 
 \def\PD      {\ensuremath{\mathit{D}}\xspace}

 \def\PJ      {\ensuremath{\mathit{J}}\xspace}
 \def\PK      {\ensuremath{\mathit{K}}\xspace}

 \def\Pb      {\ensuremath{\mathit{b}}\xspace}
 \def\Pc      {\ensuremath{\mathit{c}}\xspace}
 \def\Pd      {\ensuremath{\mathit{d}}\xspace}

 \def\Pi      {\ensuremath{\mathit{i}}\xspace}

 \def\Pp      {\ensuremath{\mathit{p}}\xspace}
 \def\Pq      {\ensuremath{\mathit{q}}\xspace}
 
 \def\Ps      {\ensuremath{\mathit{s}}\xspace}
 
 \def\Pu      {\ensuremath{\mathit{u}}\xspace}

}
{
 
 \def\Pgamma      {\ensuremath{\gamma}\xspace}

 \def\Peta        {\ensuremath{\eta}\xspace}

 \def\Pmu         {\ensuremath{\mu}\xspace}
 \def\Pnu         {\ensuremath{\nu}\xspace}
 
 \def\Ppi         {\ensuremath{\pi}\xspace}

 \def\Pchi        {\ensuremath{\chi}\xspace}
 \def\Ppsi        {\ensuremath{\psi}\xspace}
 
 \mathchardef\PDelta="7101
 \mathchardef\PXi="7104
 \mathchardef\PLambda="7103
 \mathchardef\PSigma="7106
 \mathchardef\POmega="710A
 \mathchardef\PUpsilon="7107
 
 \def\PB      {\ensuremath{B}\xspace}
 
 \def\PD      {\ensuremath{D}\xspace}

 \def\PJ      {\ensuremath{J}\xspace}
 \def\PK      {\ensuremath{K}\xspace}

 \def\Pb      {\ensuremath{b}\xspace}
 \def\Pc      {\ensuremath{c}\xspace}
 \def\Pd      {\ensuremath{d}\xspace}

 \def\Pi      {\ensuremath{i}\xspace}

 \def\Pp      {\ensuremath{p}\xspace}
 \def\Pq      {\ensuremath{q}\xspace}
 
 \def\Ps      {\ensuremath{s}\xspace}
 
 \def\Pu      {\ensuremath{u}\xspace}

}

\def\mup        {{\ensuremath{\Pmu^+}}\xspace}
\def\neu        {{\ensuremath{\Pnu}}\xspace}

\def\neum       {{\ensuremath{\neu_\mu}}\xspace}

\def\g      {{\ensuremath{\Pgamma}}\xspace}

\def\quark     {{\ensuremath{\Pq}}\xspace}

\def\uquark    {{\ensuremath{\Pu}}\xspace}

\def\dquark    {{\ensuremath{\Pd}}\xspace}

\def\squark    {{\ensuremath{\Ps}}\xspace}
\def\squarkbar {{\ensuremath{\overline \squark}}\xspace}

\def\cquark    {{\ensuremath{\Pc}}\xspace}
\def\cquarkbar {{\ensuremath{\overline \cquark}}\xspace}

\def\bquark    {{\ensuremath{\Pb}}\xspace}
\def\bquarkbar {{\ensuremath{\overline \bquark}}\xspace}

\def\pion   {{\ensuremath{\Ppi}}\xspace}

\def\pip    {{\ensuremath{\pion^+}}\xspace}
\def\pim    {{\ensuremath{\pion^-}}\xspace}

\def\kaon    {{\ensuremath{\PK}}\xspace}
  \def\Kbar    {{\kern 0.2em\overline{\kern -0.2em \PK}{}}\xspace}

\def\KorKbar    {\kern 0.18em\optbar{\kern -0.18em K}{}\xspace}

\def\Kp      {{\ensuremath{\kaon^+}}\xspace}
\def\Km      {{\ensuremath{\kaon^-}}\xspace}

  \def\Dbar    {{\kern 0.2em\overline{\kern -0.2em \PD}{}}\xspace}
\def\D       {{\ensuremath{\PD}}\xspace}

\def\DorDbar    {\kern 0.18em\optbar{\kern -0.18em D}{}\xspace}
\def\Dz      {{\ensuremath{\D^0}}\xspace}
\def\Dzb     {{\ensuremath{\Dbar{}^0}}\xspace}

\def\Dstarm  {{\ensuremath{\D^{*-}}}\xspace}

\def\Dsm     {{\ensuremath{\D^-_\squark}}\xspace}

\def\B       {{\ensuremath{\PB}}\xspace}
\def\Bbar    {{\ensuremath{\kern 0.18em\overline{\kern -0.18em \PB}{}}}\xspace}

\def\BorBbar    {\kern 0.18em\optbar{\kern -0.18em B}{}\xspace}
\def\Bz      {{\ensuremath{\B^0}}\xspace}

\def\Bu      {{\ensuremath{\B^+}}\xspace}
\def\Bps      {{\ensuremath{\B^{*+}}}\xspace}

\def\Bp      {{\ensuremath{\Bu}}\xspace}

\def\Bs      {{\ensuremath{\B^0_\squark}}\xspace}
\def\Bsss      {{\ensuremath{\B^{**0}_\squark}}\xspace}

\def\jpsi     {{\ensuremath{{\PJ\mskip -3mu/\mskip -2mu\Ppsi\mskip 2mu}}}\xspace}

\def\etac     {{\ensuremath{\Peta_\cquark}}\xspace}

\def\chicone  {{\ensuremath{\Pchi_{\cquark 1}}}\xspace}
\def\chictwo  {{\ensuremath{\Pchi_{\cquark 2}}}\xspace}
  \def\Y#1S{\ensuremath{\PUpsilon{(#1S)}}\xspace}% no space before {...}!

\def\proton      {{\ensuremath{\Pp}}\xspace}

\def\Xires       {{\ensuremath{\PXi}}\xspace}

\def\Lz          {{\ensuremath{\PLambda}}\xspace}
\def\Lbar        {{\ensuremath{\kern 0.1em\overline{\kern -0.1em\PLambda}}}\xspace}
\def\LorLbar    {\kern 0.18em\optbar{\kern -0.18em \PLambda}{}\xspace}

\def\L1520          {{\ensuremath{\PLambda(1520)}}\xspace}

\def\Lb      {{\ensuremath{\Lz^0_\bquark}}\xspace}

\def\Lc      {{\ensuremath{\Lz^+_\cquark}}\xspace}

\def\Xib     {{\ensuremath{\Xires_\bquark}}\xspace}
\def\Xibz    {{\ensuremath{{\Xires}^0_\bquark}}\xspace}
\def\Xibm    {{\ensuremath{\Xires^-_\bquark}}\xspace}

\def\Xicp    {{\ensuremath{\Xires^+_\cquark}}\xspace}

\def\BF         {{\ensuremath{\mathcal{B}}}\xspace}

\newcommand{\decay}[2]{\ensuremath{#1\!\to #2}\xspace}         % {\Pa}{\Pb \Pc}

\def\to                 {\ensuremath{\rightarrow}\xspace}

\def\Vub  {{\ensuremath{V_{\uquark\bquark}}}\xspace}
\def\Vcb  {{\ensuremath{V_{\cquark\bquark}}}\xspace}

\def\bsll     {\decay{\bquark}{\squark \ell^+ \ell^-}}

\def\AT#1     {\ensuremath{A_{\mathit{T}}^{#1}}\xspace}           % 2

\def\C#1      {\ensuremath{\mathcal{C}_{#1}}\xspace}                       % 9
\def\Cp#1     {\ensuremath{\mathcal{C}_{#1}^{'}}\xspace}                    % 7
\def\Ceff#1   {\ensuremath{\mathcal{C}_{#1}^{\mathit{(eff)}}}\xspace}        % 9
\def\Cpeff#1  {\ensuremath{\mathcal{C}_{#1}^{'\mathit{(eff)}}}\xspace}       % 7
\def\Ope#1    {\ensuremath{\mathcal{O}_{#1}}\xspace}                       % 2
\def\Opep#1   {\ensuremath{\mathcal{O}_{#1}^{'}}\xspace}                    % 7

\newcommand{\tev}{\ifthenelse{\boolean{inbibliography}}{\ensuremath{~T\kern -0.05em eV}}{\ensuremath{\mathit{\,Te\kern -0.1em V}}}\xspace}
\newcommand{\gev}{\ensuremath{\mathrm{\,Ge\kern -0.1em V}}\xspace}
\newcommand{\mev}{\ensuremath{\mathrm{\,Me\kern -0.1em V}}\xspace}
\newcommand{\kev}{\ensuremath{\mathrm{\,ke\kern -0.1em V}}\xspace}
\newcommand{\ev}{\ensuremath{\mathrm{\,e\kern -0.1em V}}\xspace}
\newcommand{\gevc}{\ensuremath{{\mathrm{\,Ge\kern -0.1em V\!/}c}}\xspace}
\newcommand{\mevc}{\ensuremath{{\mathrm{\,Me\kern -0.1em V\!/}c}}\xspace}
\newcommand{\gevcc}{\ensuremath{{\mathrm{\,Ge\kern -0.1em V\!/}c^2}}\xspace}
\newcommand{\gevgevcccc}{\ensuremath{{\mathrm{\,Ge\kern -0.1em V^2\!/}c^4}}\xspace}
\newcommand{\mevcc}{\ensuremath{{\mathrm{\,Me\kern -0.1em V\!/}c^2}}\xspace}

\def\invfb   {\ensuremath{\mbox{\,fb}^{-1}}\xspace}

\def\fs   {\ensuremath{\mathrm{ \,fs}}\xspace}

\def\gsim{{~\raise.15em\hbox{$>$}\kern-.85em
          \lower.35em\hbox{$\sim$}~}\xspace}
\def\lsim{{~\raise.15em\hbox{$<$}\kern-.85em
          \lower.35em\hbox{$\sim$}~}\xspace}

\def\sqs   {\ensuremath{\protect\sqrt{s}}\xspace}

\def\pt         {\mbox{$p_{\mathrm{ T}}$}\xspace}

\def\tell1  {TELL1\xspace}
\def\ukl1   {UKL1\xspace}

\def\Xibc{{\ensuremath{{\Xires}_{\bquark\cquark}^{0}}}\xspace}
\def\Omegabc{{\ensuremath{{\Omega}_{\bquark\cquark}^{0}}}\xspace}

\usepackage[bitstream-charter]{mathdesign}
\DeclareSymbolFont{usualmathcal}{OMS}{cmsy}{m}{n}
\DeclareSymbolFontAlphabet{\mathcal}{usualmathcal}

\begin{document}

\begin{center}{\Large \textbf{
Studies of \bquark-hadrons and quarkonia at LHCb \\
}}\end{center}

% TODO: write the author list here. Use initials + surname format.
% Separate subsequent authors by a comma, omit comma at the end of the list.
% Mark the corresponding author with a superscript *.
\begin{center}
Zhiyu Xiang\textsuperscript{}
%Aah B. Cee\textsuperscript{2} and
%Gee K. See\textsuperscript{3$\star$}
\end{center}

\begin{center}
{\bf } University of Chinese Academy of Sciences, Beijing, China
\\
%{\bf 2} Affiliation2
%\\
%{\bf 3} Affiliation2
%\\
zxiang@cern.ch
\end{center}

\begin{center}
\today
\end{center}

% For convenience during refereeing (optional),
% you can turn on line numbers by uncommenting the next line:
%\linenumbers

\definecolor{palegray}{gray}{0.95}
\begin{center}
\colorbox{palegray}{
  \begin{tabular}{rr}
  \begin{minipage}{0.1\textwidth}
    \includegraphics[width=22mm]{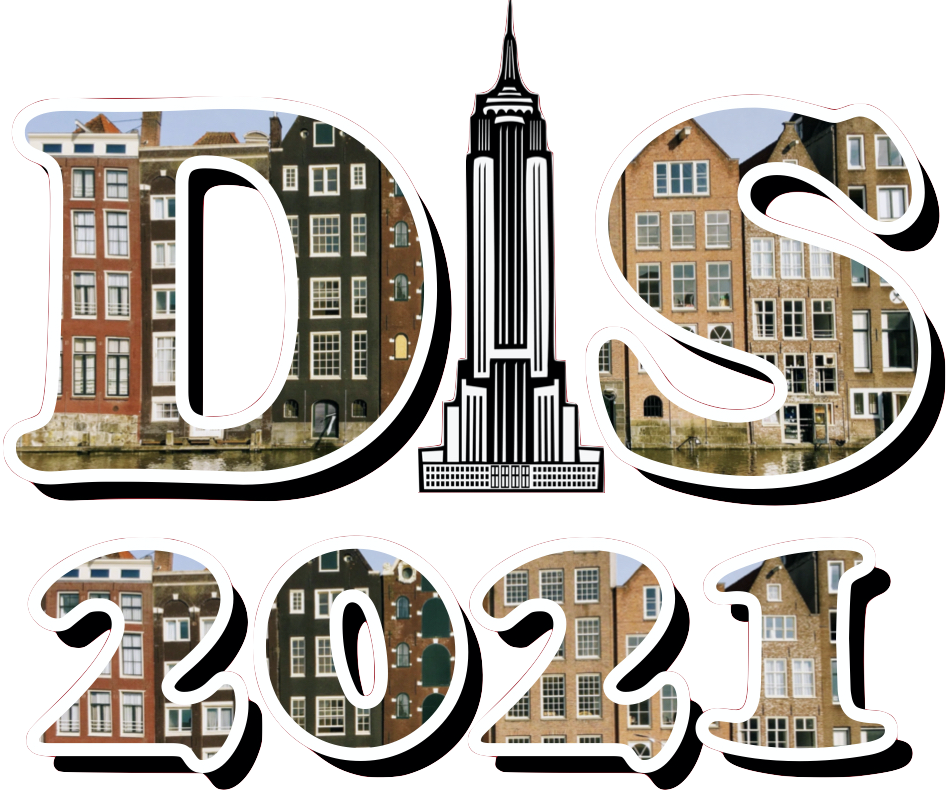}
  \end{minipage}
  &
  \begin{minipage}{0.75\textwidth}
    \begin{center}
    {\it Proceedings for the XXVIII International Workshop\\ on Deep-Inelastic Scattering and
Related Subjects,}\\
    {\it Stony Brook University, New York, USA, 12-16 April 2021} \\
    \doi{10.21468/SciPostPhysProc.?}\\
    \end{center}
  \end{minipage}
\end{tabular}
}
\end{center}

\section*{Abstract}
{\bf
% TODO: write your abstract here.
LHCb has productive studies of quarkonia and \bquark-hadrons which provide essential inputs to study quantum chromodynamics (QCD). In this proceeding, seven recent results are reported: the observation of a new excited \Xibz resonance in \Xibm\pip final states, the new excited \Bs state is observed in \Bp\Km mass spectrum, the first observation of suppressed semileptonic \Bs\to\Km\mup\neum decay, the first observation of the decay \Bz\to\Dz\Dzb\Kp\pim, the first observation of the Cabibbo-suppressed decay \Lb\to\chicone\proton\pim, the first observation of \Lb\to\Lc\Kp\Km\pim decay, and end this report with the first search for the doubly heavy baryon \Omegabc and a search for \Xibc baryon.
}

%\vspace{10pt}
%\noindent\rule{\textwidth}{1pt}
%\tableofcontents\thispagestyle{fancy}
%\noindent\rule{\textwidth}{1pt}
%\vspace{10pt}

\section{Introduction}
% TODO: write your article here.
Heavy flavour hadrons are among the most important tools for the study of QCD in high-energy hadronic collisions. At the LHC, \bquark-hadrons production is dominated by the gluon-fusion heavy quark pair production subprocess and can be predicted by considering the partonic cross section for heavy quark pair production with input parton distribution function and fragmentation functions. In addition, many aspects of \bquark-decays can be understood through the Heavy Quark Effective Theory (HQET)~\cite{rf1}. So the study of \bquark-hadrons provides a crucial testing ground for perturbative and nonperturbative QCD. 
The LHCb detector~\cite{rf2,rf3}, with its high momentum resolution, great particle identification capability and flexible trigger strategy, is perfectly suited to study  features.

\section{Observation of a new \Xibz state}
The \Xibz and \Xibm states form an isodoublet of \bquark\squark\quark bound states is allowed in the constituent quark model~\cite{rf4}, where \quark is a \uquark or \dquark quark, respectively. Three such isodoublets are predicted which have different $j_{qs}$ and $J^P$ quantum numbers, where $j_{qs}$ is the spin of the light diquark system \quark\squark, and  $J^P$ represents the spin and parity of the baryon state. Beyond these lowest-lying states, a spectrum of heavier states is also expected. The \Xibm excitation with a mass of 6227 \mevcc have been observed in the \Lb\Km and \Xibz\pim invariant-mass spectra, which is consistent with a $P$-wave excitation in subsequent constituent quark model but could also be wholly or partially molecular in some other investigations. More information on the observed states, or observation of
additional excited beauty-baryon states, will provide additional input for these theoretical works.

Using $pp$ collision data with an integrated luminosity of 8.5 \invfb recorded by the LHCb at $\sqs=7,8 \text{\ and } 13\ \rm{TeV}$, a new \Xibz state, mass value around 6227 \mevcc, is seen through its decay to the \Xibm\pip final state~\cite{rf5}. The mass and natural width of $\Xib(6227)^0$ are measured to be
\begin{equation*}
\begin{aligned}
m(\Xib(6227)^0)&=6227.1^{+1.4}_{-1.5}\pm0.5\ \mevcc,\\
\Gamma(\Xib(6227)^0)&=18.6^{+5.0}_{-4.1}\pm1.4\ \mevcc.\\
\end{aligned}
\end{equation*}
where the first uncertainty is statistical and the second is experimental systematic.
The relative production rate of the $\Xib(6227)^0$ state at $\sqs=13\ \rm{TeV}$ is measured through its decay to \Xibm\pip and found to be consistent with that of $\Xib(6227)^-\to\Xibz\pim$ decay.

\section{Observation of new excited \Bs states}
Potential models exploiting heavy-quark symmetry~\cite{rf6} are used to calculate properties of mesons containing one heavy and one light quark, including those with \bquarkbar\squark quark content, termed collectively the mesons \Bsss mesons. Yet it is still difficult to predict precise masses and widths of such states. Experimental measurements are therefore essential.

\begin{figure}[!h]
\centering
\includegraphics[width=0.45\textwidth]{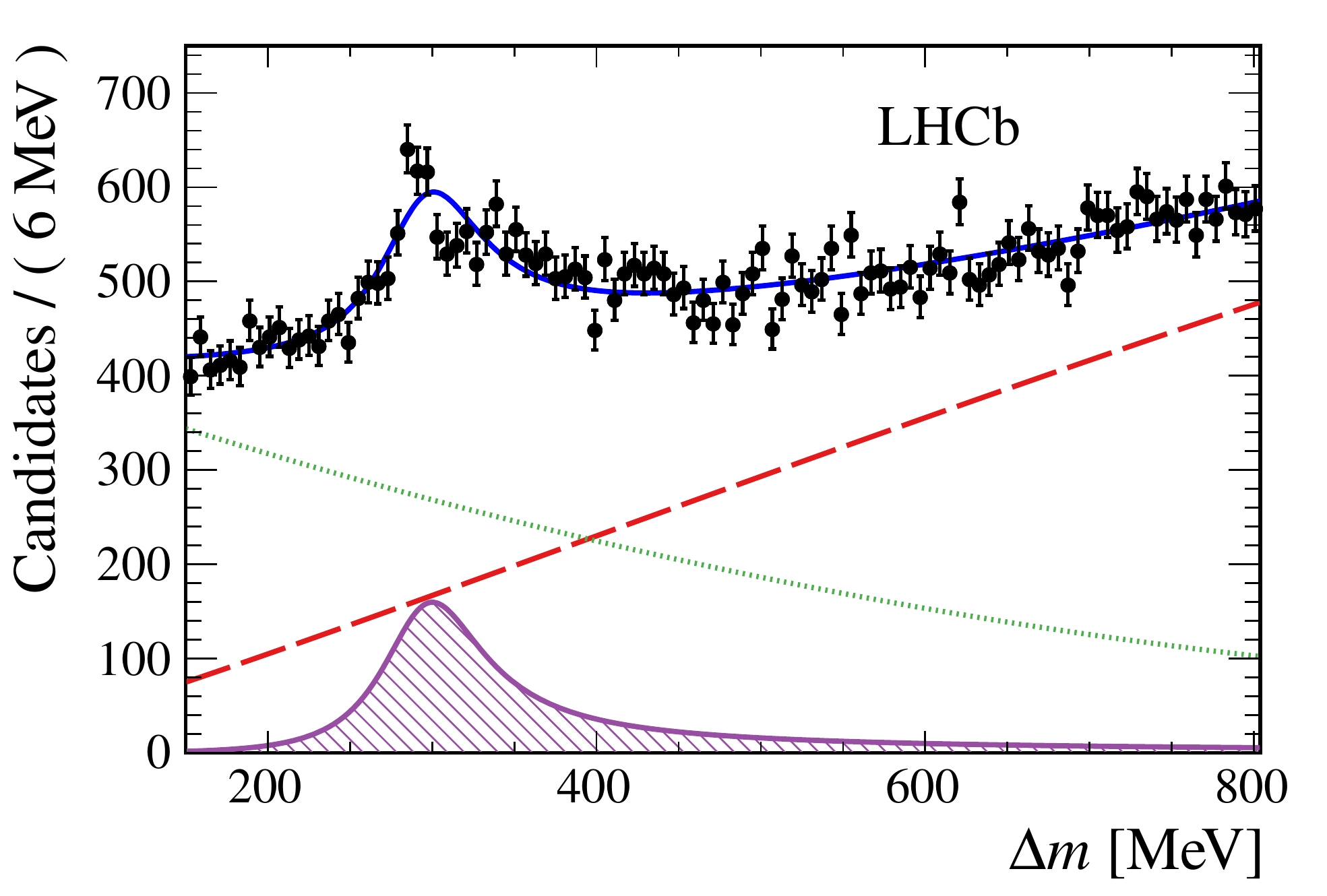}
\includegraphics[width=0.45\textwidth]{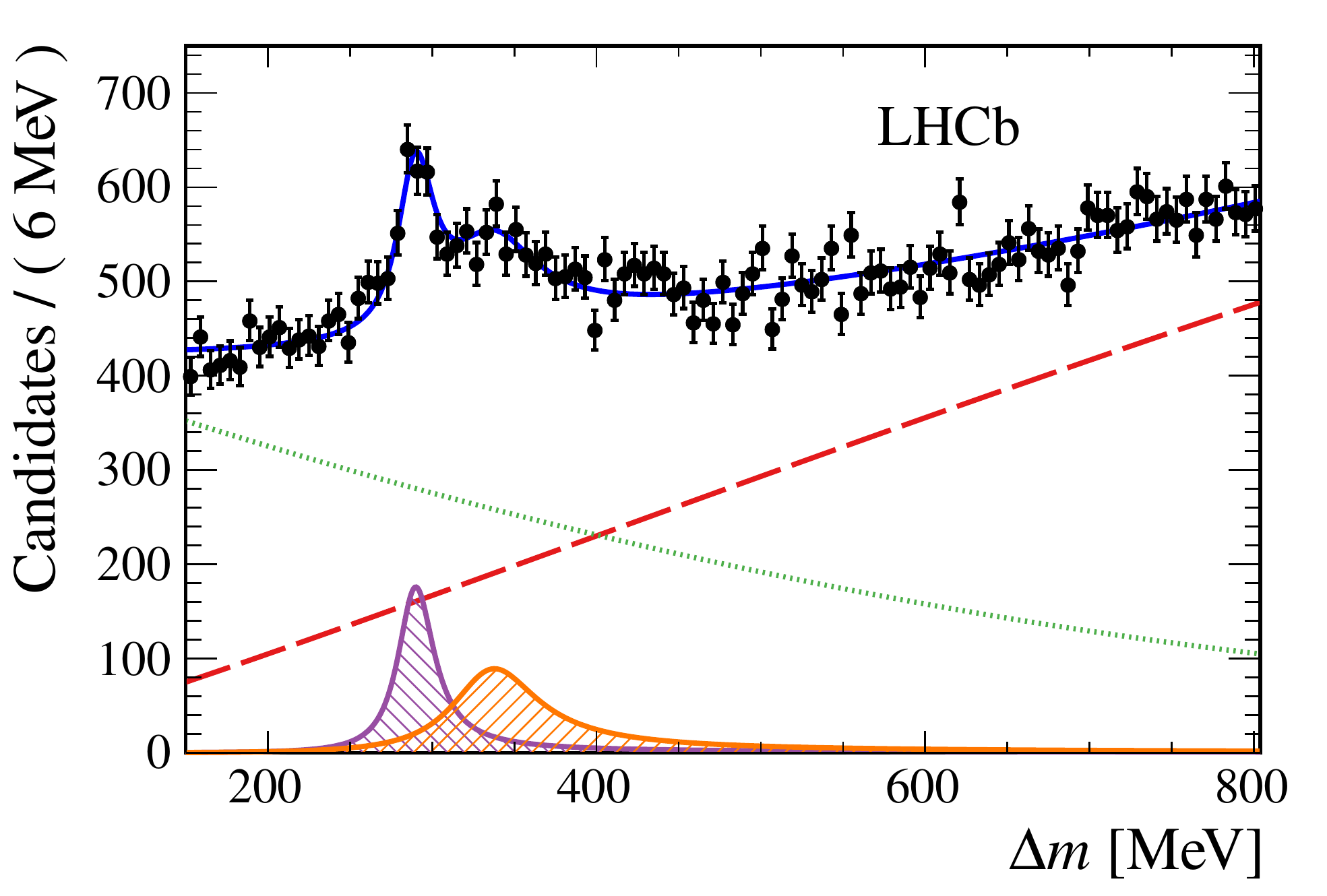}
\caption{For candidates with the prompt kaon $\pt>2\gev$, the \Bp\Km mass difference distributions in data, overlaid with the fit models: (left plot) one-peak hypothesis and (right plot) two-peak hypothesis.}
\label{fig1}
\end{figure}

LHCb observed a structure in the \Bp\Km invariant mass spectrum in a data sample of $pp$ collisions at \mbox{$\sqs=7,8 \text{\ and } 13\ \rm{TeV}$}, corresponding to a total integrated luminosity of 9 \invfb~\cite{rf7}. Take the candidates with the prompt kaon $\pt>2\gev$ as an example. An obvious structure appears in the mass difference $\Delta m$ around 300\mevcc, where $\Delta m=m_{\Bp\Km}-m_{\Bp}-m_{\Km}$, as shown in Figure~\ref{fig1}. The fit results compare first the one-peak hypothesis to the null hypothesis, and then the two-peak to the one-peak hypothesis. The minimum of the test statistics across each fit with different associated production descriptions gives local significances of more than 20$\sigma$ for the one-peak fit with respect to the null hypothesis and 7.7$\sigma$ for the two-peak description with respect to the one-peak hypothesis.

A single resonance which can decay through both \Bp\Km and \Bps\Km is disfavoured but can't be excluded. If consider the decay is directly to \Bp\Km, the measured parameters are
\begin{equation*}
\begin{aligned}
m_{1} &=6063.5 \pm 1.2 \text { (stat) } \pm 0.8 \text { (syst) } \mathrm{\mevcc},\\
\Gamma_{1} &=26 \pm 4 \text { (stat) } \pm 4 \text { (syst) } \mathrm{\mevcc}, \\
m_{2} &=6114 \pm 3(\text { stat }) \pm 5 \text { (syst) } \mathrm{\mevcc}, \\
\Gamma_{2} &=66 \pm 18(\text { stat }) \pm 21 \text { (syst) } \mathrm{\mevcc}.
\end{aligned} 
\end{equation*}

\section{Observation of \Bs\to\Km\mup\neum and a measurement of $|\Vub|/|\Vcb|$}
The precision measurement of the Cabibbo-Kobayashi-Maskawa (CKM) matrix elements could be a probe on physics beyond Standard Model. \bquark-hadrons can decay weakly via \mbox{$\bquark\to\cquark (W^*\to\ell\neu)$} and \mbox{$\bquark\to\uquark (W^*\to\ell\neu)$}. The involved \Vub and \Vcb obey
the observed hierarchy $|\Vub|/|\Vcb|\sim0.1$. Their exclusive measurements show a discrepancy with inclusive results.

\Bs\to\Km\mup\neum is observed firstly using $pp$ collision data collected by LHCb at $\sqs=8\ \rm{TeV}$, with a total integrated luminosity of 2 \invfb~\cite{rf8}. To measure $|\Vub|/|\Vcb|$ and reduce the systematics on its branching fraction, the \Bs\to\Dsm\mup\neum is taken as normalization
channel. The measurement is performed in two \Bs\to\Km momentum transfer regions, $q^{2}<7\gevgevcccc$ (low) and $q^{2}>7\gevgevcccc$ (high). To discriminate signal and different backgrounds, The \Bs mass is represented by the corrected mass, defined as $m_{\text {corr }}=\sqrt{m_{Y \mu}^{2}+p_{\perp}^{2} / c^{2}}+p_{\perp} / c$, where $Y=\Km\ \text{or}\ \Dsm$ and $p_{\perp}$ is the
momentum of this pair transverse to the \Bs flight direction.

\begin{figure}[!h]
\centering
\includegraphics[width=0.45\textwidth]{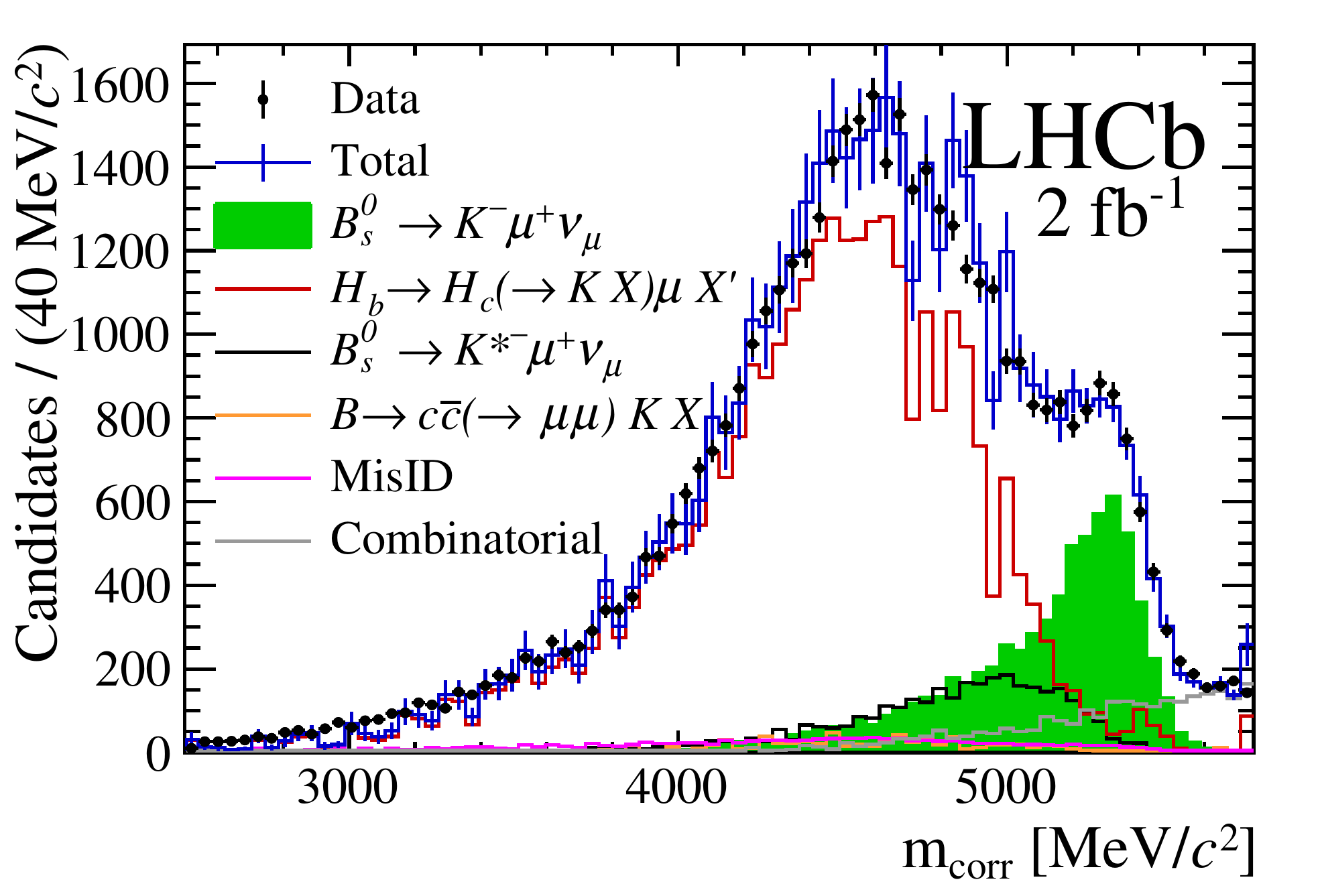}
\includegraphics[width=0.45\textwidth]{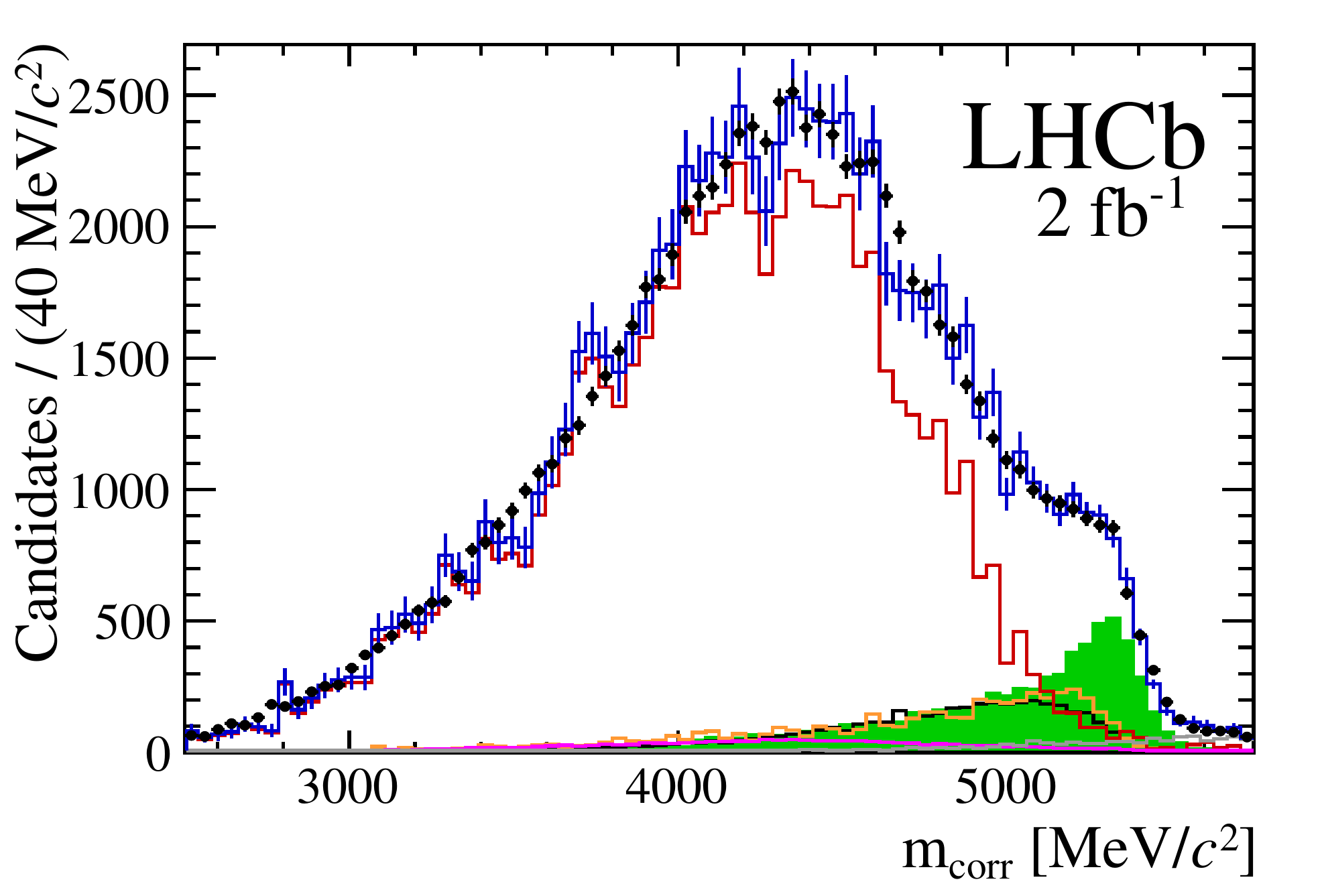}
\caption{Distribution of $m_{\text {corr }}$ for the signal \Bs\to\Km\mup\neum with (low) $q^{2}<7\gevgevcccc$ and (high) $q^{2}>7\gevgevcccc$ region.}
\label{fig2}
\end{figure}

As shown in Figure~\ref{fig2}, the $m_{\text {corr }}$ fit template for the signal is obtained from simulation, while the shapes for the background components are derived from either simulation or control samples.
With obtained about 13 thousand \Bs\to\Km\mup\neum candidates and about 200 thousand \Bs\to\Dsm\mup\neum candidates, the ratio of branching
fractions between signal and normalisation channels is determined to be

\begin{equation*}
\begin{aligned}
R_{\mathrm{BF}}(\text { low }) &=\left(1.66 \pm 0.08(\text { stat }) \pm 0.07 \text { (syst) } \pm 0.05\left(D_{s}\right)\right) \times 10^{-3}, \\
R_{\mathrm{BF}}(\text { high }) &=\left(3.25 \pm 0.21(\text { stat })_{-0.17}^{+0.16}(\text { syst }) \pm 0.09\left(D_{s}\right)\right) \times 10^{-3}, \\
R_{\mathrm{BF}}(\text { all }) &=\left(4.89 \pm 0.21(\text { stat })_{-0.21}^{+0.20}(\text { syst }) \pm 0.14\left(D_{s}\right)\right) \times 10^{-3}.
\end{aligned}
\end{equation*}
The ratio of CKM elements $|\Vub|/|\Vcb|$ is measured to be
\begin{equation*}
\begin{aligned}
\left|V_{u b}\right| /\left|V_{c b}\right|(\text { low }) &=0.0607 \pm 0.0015 \text { (stat) } \pm 0.0013 \text { (syst) } \pm 0.0008\left(D_{s}\right) \pm 0.0030(\mathrm{FF}), \\
\left|V_{u b}\right| /\left|V_{c b}\right|(\text { high }) &=0.0946 \pm 0.0030 \text { (stat) }_{-0.0025}^{+0.0024}(\text { syst }) \pm 0.0013\left(D_{s}\right) \pm 0.0068(\mathrm{FF}),
\end{aligned}
\end{equation*}
where the last term refer to the form
factor integrals. The discrepancy of obtained $|\Vub|/|\Vcb|$ between low and high $q^2$ region related to the difference in the theoretical calculations of the form factors.

\section{First observation of the decay \Bz\to\Dz\Dzb\Kp\pim}
The \bquark\to\cquark\cquarkbar\squark transitions involved $\B\to\D^{(*)}\bar{D}^{(*)}K$ decay offering the opportunity to search for new \cquark\squarkbar or \cquark\cquarkbar states. These processes can also provide important information to calculations of the \cquark\cquarkbar contribution above the open-charm threshold in \bsll decays. Yet the decay involving $\B\to\D^{(*)}\bar{D}^{(*)}K\pi$ transitions have not been observed.

\begin{figure}[!h]
\centering
\includegraphics[width=0.45\textwidth]{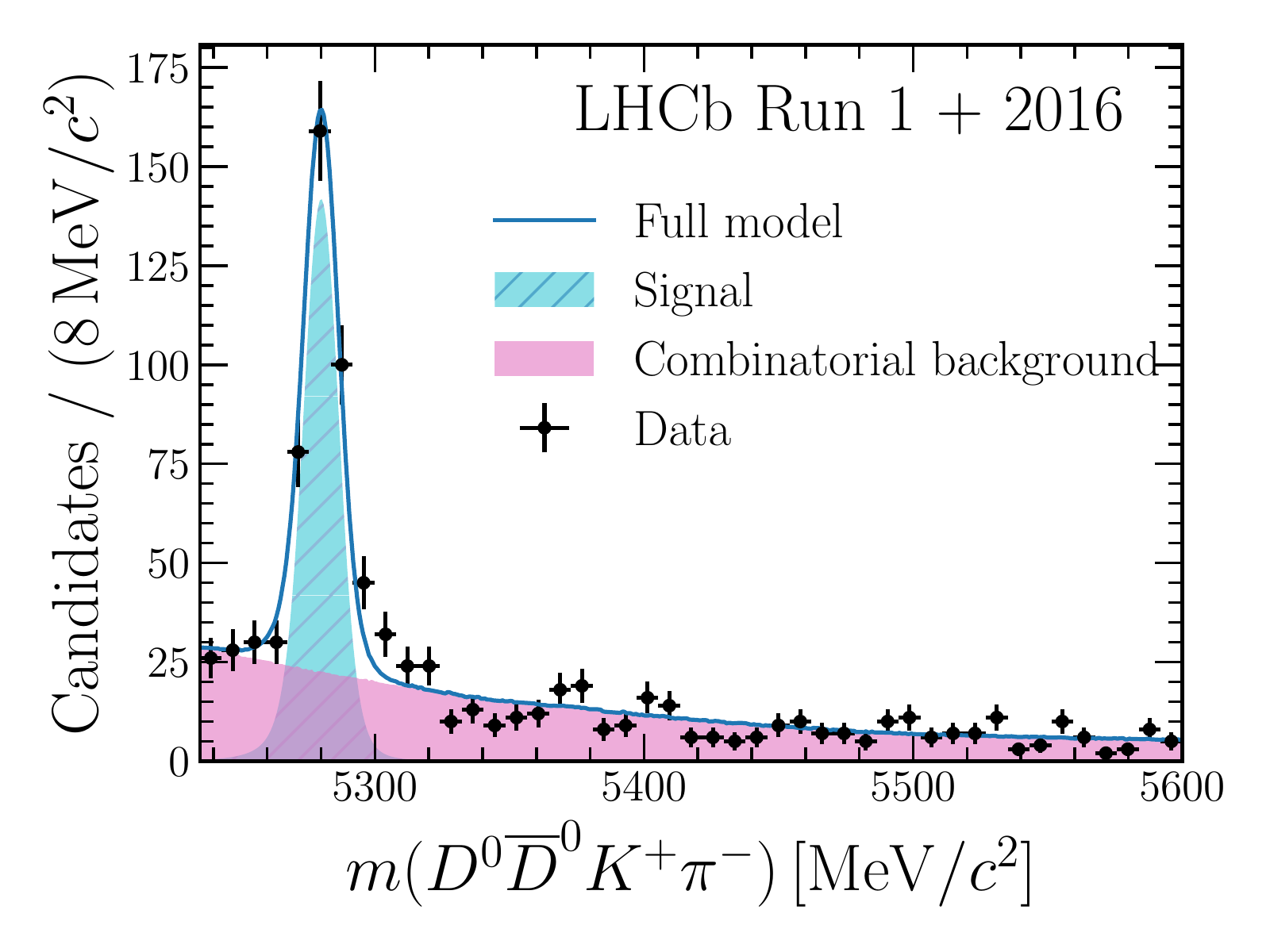}
\includegraphics[width=0.45\textwidth]{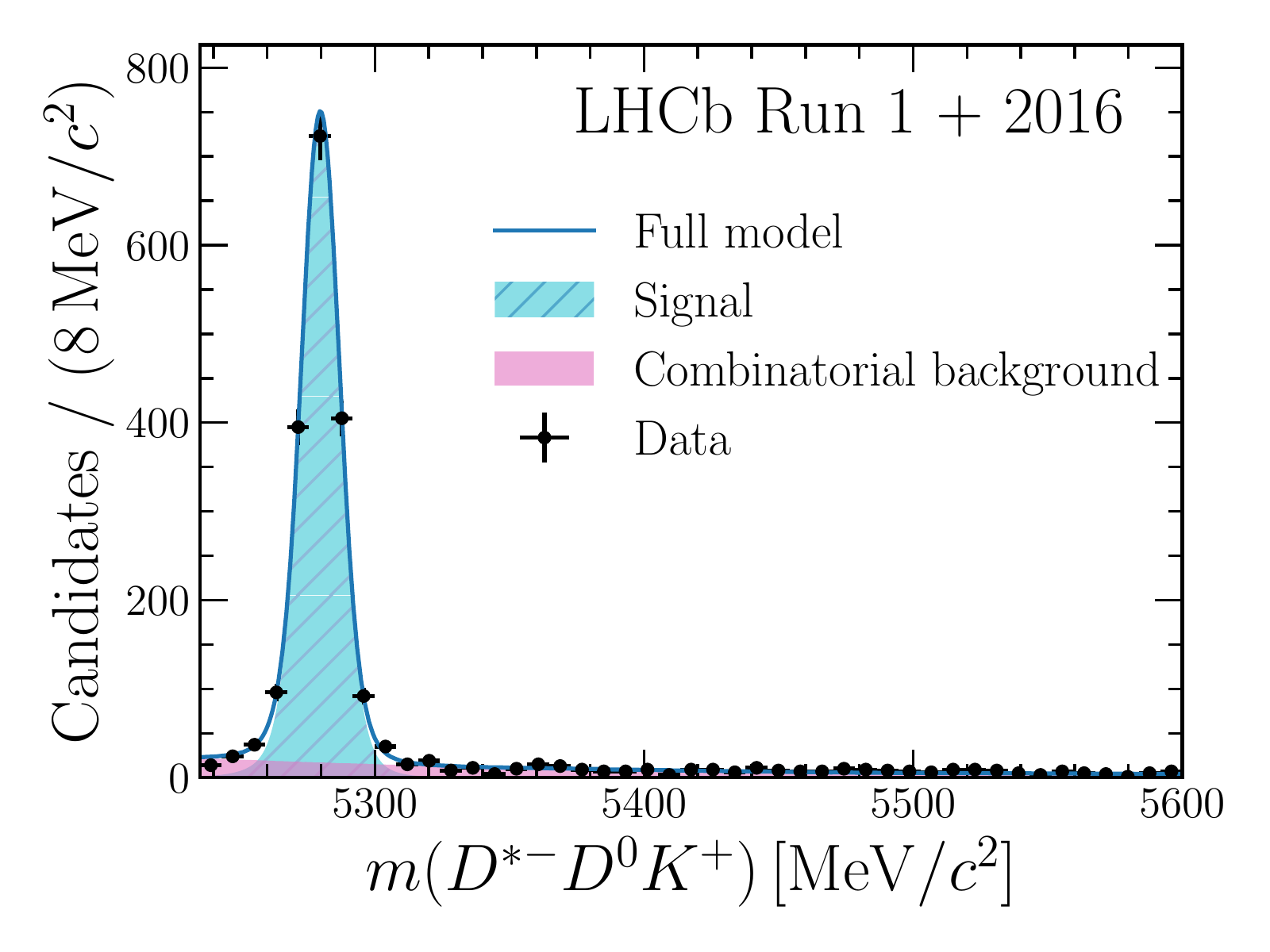}
\caption{Invariant-mass distributions and fit projections for \Bz candidates in (left) the signal and (right) control mode for all subsamples combined. The small single-charm and charmless background is included in the signal component.}
\label{fig3}
\end{figure}

Recently, the \Bz\to\Dz\Dzb\Kp\pim decay, excluding contributions from \Bz\to\Dstarm\Dz\Kp transitions, first observed using $pp$ collision data at $\sqs=7,8 \text{\ and } 13\ \rm{TeV}$ collected by LHCb~\cite{rf9}, corresponding to an integrated luminosity of 4.7\invfb. The fit results on mass spectra are shown in Figure~\ref{fig3}. The normalization mode \Bz\to\Dstarm\Dz\Kp is excluded in signal range by the requirement of \mbox{$|m(\Dzb \pim)-m(\Dzb)-[m_{\rm{PDG}}(\Dstarm)-m_{\rm{PDG}}(\Dzb)]|>(4 \times 0.724) \mevcc$}. The absolute branching fraction of \Bz\to\Dz\Dzb\Kp\pim is determined to be $(3.50 \pm 0.27 \pm 0.26 \pm 0.30) \times 10^{-4}$, where the first uncertainty is statistical, the second is systematic and the third is related to the uncertainties in the branching fractions of the \Bz\to\Dstarm\Dz\Kp.

\section{Observation of the decay \Lb\to\chicone\proton\pim}
\begin{figure}[!h]
\centering
\includegraphics[width=0.44\textwidth]{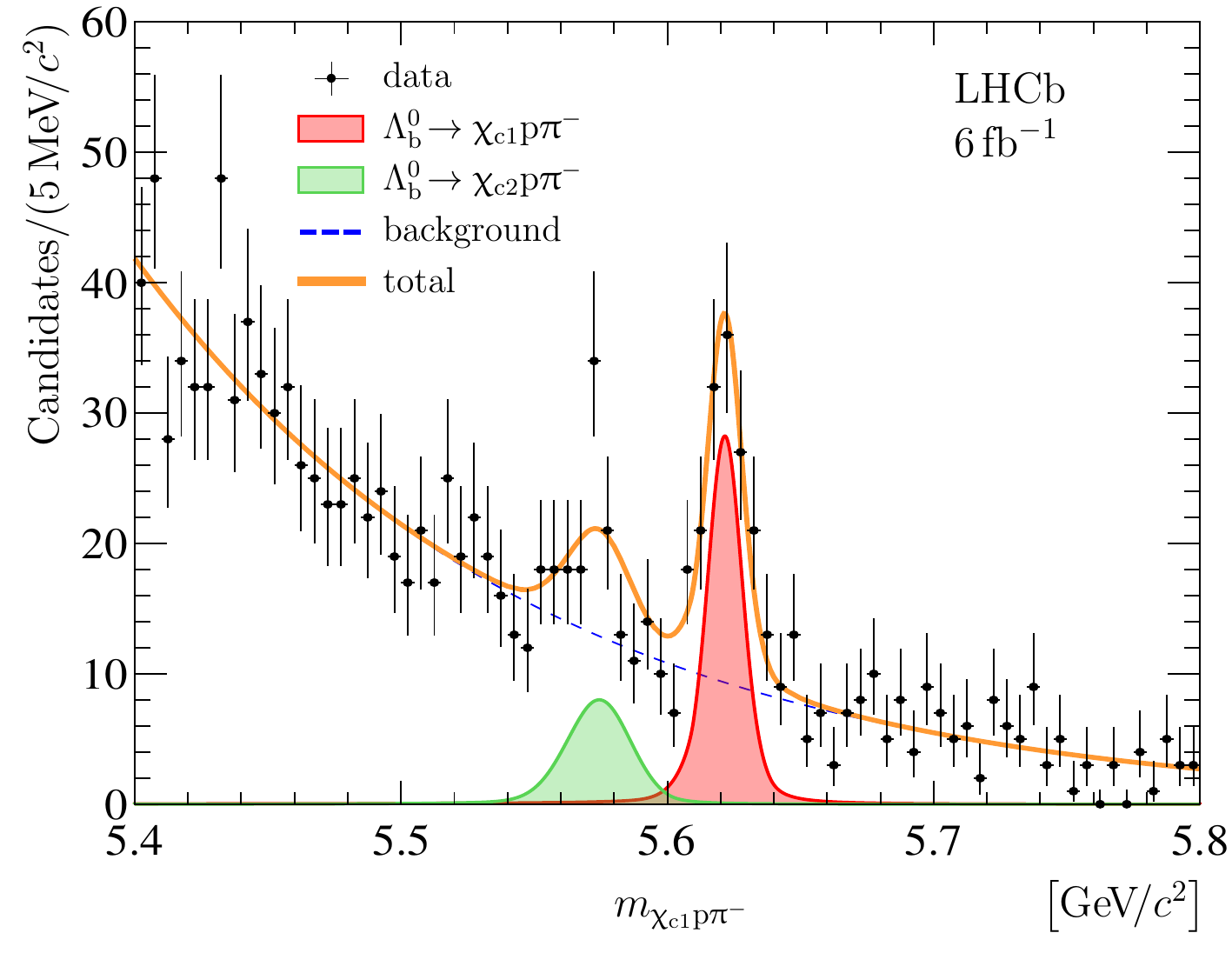}
\includegraphics[width=0.46\textwidth]{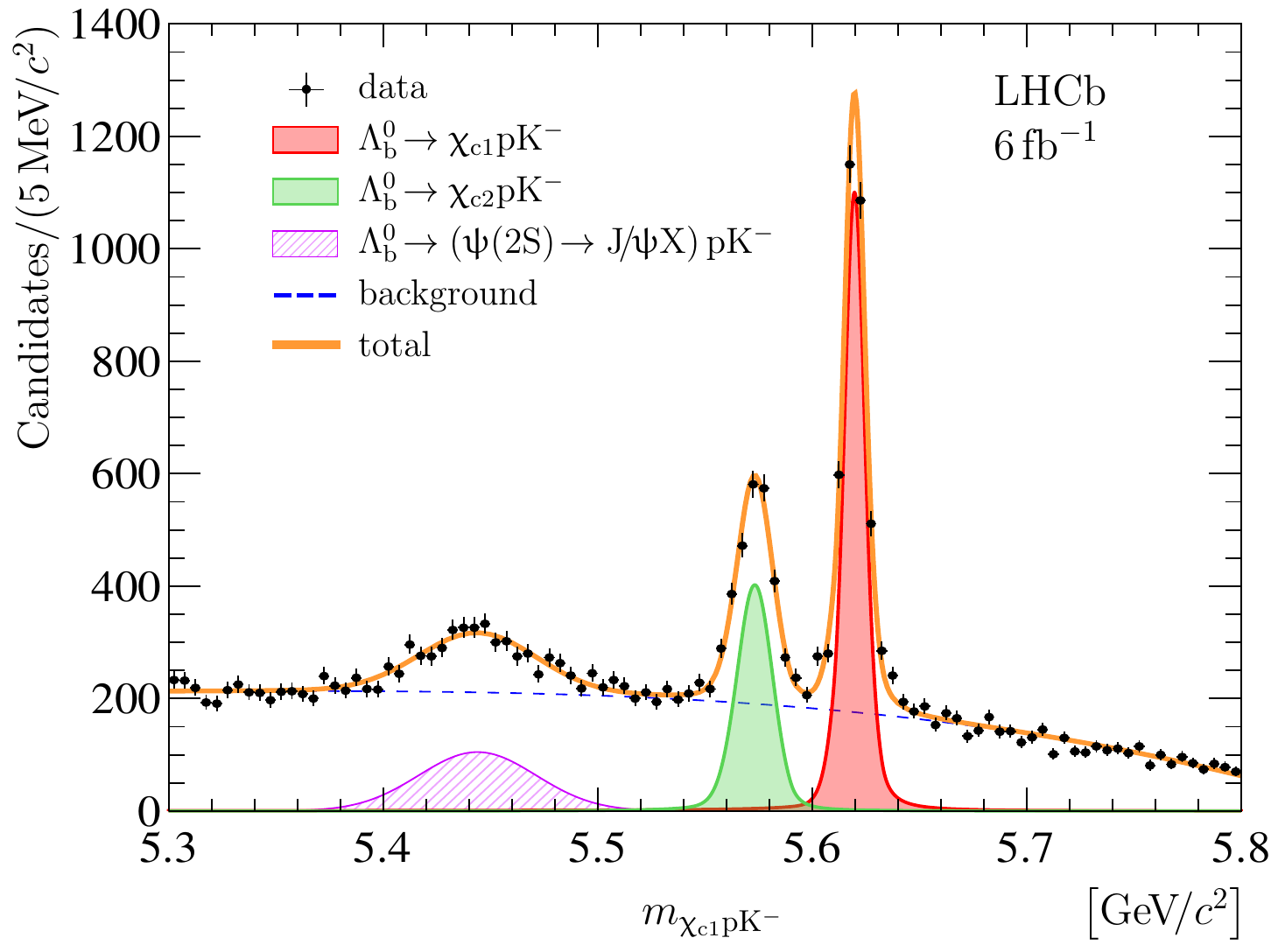}
\caption{Mass distribution for selected \Lb\to\chicone\proton\pim (left) and \Lb\to\chicone\proton\Km (right).}
\label{fig4}
\end{figure}

The hidden-charm pentaquark resonances have been observed only in the \jpsi\proton, \jpsi\Lz systems. It also possible hidden in other decay modes, such as $\etac\proton,\chicone\proton$ and \chictwo\proton systems. Using $pp$ collision data with an integrated luminosity of 6 \invfb recorded by the LHCb at $\sqs=13\ \rm{TeV}$, a search for the \Lb\to\chicone\proton\pim and \Lb\to\chictwo\proton\pim is performed~\cite{rf10}. The \chicone and \chictwo mesons are reconstructed via \jpsi\g final states. The \Lb\to\chicone\proton\pim decay is observed for the first time and the \Lb\to\chictwo\proton\pim  decay is obtained with a significance of $3.5\sigma$, as shwon in Figure~\ref{fig4}. With the current statistics of \Lb\to\chicone\proton\pim decay, no evidence for contributions from exotic states is found. The ratios of the branching fractions are measured to be
\begin{equation*}
\begin{aligned}
R_{\pion/\kaon}&=\frac{\BF(\Lb\to\chicone\proton\pim)}{\BF(\Lb\to\chicone\proton\Km)}=(6.59 \pm 1.01 \pm 0.22) \times 10^{-2}, \\
R^{\pion}_{2/1}&=\frac{\BF(\Lb\to\chictwo\proton\pim)}{\BF(\Lb\to\chicone\proton\pim)}=0.95 \pm 0.30 \pm 0.04 \pm 0.04, \\
R^{\kaon}_{2/1}&=\frac{\BF(\Lb\to\chictwo\proton\Km)}{\BF(\Lb\to\chicone\proton\Km)}=1.06 \pm 0.05 \pm 0.04 \pm 0.04.
\end{aligned}
\end{equation*}
where the first uncertainty is statistical, the second is systematic and the third is related to the uncertainties in the branching fractions of the $\chi_{c1,2}\to\jpsi\g$.

\section{Observation of the \Lb\to\Lc\Kp\Km\pim decay}
Many \bquark-baryon decays have not been observed or poorly measured. The \Lb\to\Lc\Kp\Km\pim decay is observed for the first time by LHCb using a data sample of $pp$ collisions at $\sqs=7,8\ \rm{TeV}$, corresponding to a total integrated luminosity of 3 \invfb~\cite{rf11}. This decay provides a laboratory to search for open-charm pentaquarks with valence quark content \cquark\squarkbar\uquark\uquark\dquark that could decay strongly to the \Lc\Kp final state. With obtained signal statistics, $3400\pm80$, no structure is observed in \Lc\Kp mass spectrum. The branching fraction is determined,  using the \Lb\to\Lc\Dsm decay as a normalisation channel, to be
\begin{equation*}
\BF(\Lb\to\Lc\Kp\Km\pim)=(1.02 \pm 0.03 \pm 0.05 \pm 0.10) \times 10^{-3},
\end{equation*}
where the first uncertainty is statistical, the second is systematic, and the third is due to the knowledge of the \Lb\to\Lc\Dsm branching fraction.

\section{Search for the doubly heavy baryons \Omegabc and \Xibc decaying to \Lc\pim and \Xicp\pim}
The baryons with two heavy quarks are of interest for the theoretical consideration, because in this system the interaction of light quark with the heavy quarks is essential as well as the interaction between the heavy quarks.  A first search for the doubly-heavy \Xibc baryon using its decay to the \Dz\proton\Km final state is performed by LHCb~\cite{rf12}, no significant signal is found.

In \Lc\pim or \Xicp\pim final states, a new search for the \Xibc and the first search for the \Omegabc are performed using $pp$ collision data collected by LHCb at $\sqs=13\ \rm{TeV}$~\cite{rf13}, corresponding to an integrated luminosity of 5.2 \invfb. No significant excess is found in the invariant mass range from 6.7 to 7.3 \gevcc. Upper limits are set at 95\% confidence level on the ratio of the \Xibc and \Omegabc production cross-section times its branching fraction to \Lc\pim (\Xicp\pim) relative to that of the \Lb (\Xibz) baryon for different lifetime hypotheses in range of $0.2-0.4\fs$. The upper limits change from $0.5\times10^{-4}$ to $2.5\times10^{-4}$ for the \Omegabc\to\Lc\pim (\Xibc\to\Lc\pim) decay, and from $1.4\times10^{-3}$ to $6.9\times10^{-3}$ for the \Omegabc\to\Xicp\pim (\Xibc\to\Xicp\pim) decay. 

\section{Conclusion}
Several recent LHCb results on quarkonia and \bquark-hadrons in $pp$ collisions have been reported. The observation of a new excited state $\Xib(6227)^0$. The new excited \Bs state is observed in \Bp\Km final state. The first observation of suppressed semileptonic \Bs\to\Km\mup\neum decay and a measurement of $|\Vub|/|\Vcb|$. The first observation of the decay \Bz\to\Dz\Dzb\Kp\pim. The first observation of the decay \Lb\to\chicone\proton\pim. The first observation of \Lb\to\Lc\Kp\Km\pim decay. The upper limit set for the doubly heavy baryon \Omegabc and \Xibc. More results of quarkonia and \bquark-hadrons at LHCb are expected in the near future.

%%%%%%%%%%%%
% TODO:
\bibliography{SciPost_LaTeX_Template.bib}

\end{document}